\documentclass[aps,prl,twocolumn,superscriptaddress]{revtex4-1}
\usepackage{graphicx}
\usepackage{latexsym}
\usepackage{amsbsy,amsmath,amssymb}

\begin{document}

\title{Topological excitations in a Kagom\'e magnet}

\author{Manuel Pereiro}
\email[Corresponding author. ]{manuel.pereiro@physics.uu.se}
\affiliation{Department of Physics and Astronomy, Uppsala University, P.O. Box 516, 751 20 Uppsala, Sweden}
\author{Dmitry Yudin}
\affiliation{Department of Physics and Astronomy, Uppsala University, P.O. Box 516, 751 20 Uppsala, Sweden}
\author{Jonathan Chico}
\affiliation{Department of Physics and Astronomy, Uppsala University, P.O. Box 516, 751 20 Uppsala, Sweden}
\author{Corina Etz}
\affiliation{Department of Physics and Astronomy, Uppsala University, P.O. Box 516, 751 20 Uppsala, Sweden}
\author{Olle Eriksson}
\affiliation{Department of Physics and Astronomy, Uppsala University, P.O. Box 516, 751 20 Uppsala, Sweden}
\author{Anders Bergman}
\affiliation{Department of Physics and Astronomy, Uppsala University, P.O. Box 516, 751 20 Uppsala, Sweden}


\begin{abstract}
It is shown here that a Kagom\'e magnet, with Heisenberg and Dzyaloshinskii-Moriya interactions causes non trivial topological and chiral magnetic properties. Chirality---that is, left or right handedness---is a very important concept in a broad range of scientific areas, and particularly, in condensed matter physics. Inversion symmetry breaking relates chirality with skyrmions, that are protected field configurations with particle-like and topological properties. Here, the reported numerical simulations and theoretical considerations reveal that the magnetic excitations of the Kagom\'e magnet can both be of regular bulk magnon character, as well as, having a non-trivial topological nature. We also find that under special circumstances, skyrmions emerge as excitations, having stability even at room temperature. Chiral magnonic edge states of a Kagom\'e magnet offer, in addition, a promising way to create, control and manipulate skyrmions. This has potential for applications in spintronics, magnonics and skyrmionics, i.e., for information storage or as logic devices based on the transportation and control of these particles. Collisions between these particle-like excitations are found to be elastic in the skyrmion-skyrmion channel, albeit without mass-conservation for an individual skyrmion. Skyrmion-antiskyrmion collisions are found to be more complex, where annihilation and creation of these objects have a distinct non-local nature.
\end{abstract}


\maketitle

Systematic studies of strongly interacting electronic systems had been based on the formalism of Landau Fermi liquid for a long time. Landau pointed out that a strongly correlated problem can be replaced by a set of quasiparticles and effective amplitudes describing the interaction among them. Each quasiparticle is characterized by its own effective mass and can be adiabatically connected to a weakly interacting Fermi gas. A proper many-body description of solid state systems can not be addressed within linear theory, however small anharmonic perturbations do not change qualitatively the quasiparticle picture and can be treated as scattering processes between them, leading to multiple harmonic generation, quasiparticles dressing, etc\cite{Abrikosov:1975uw}. The inability to explain a variety of phenomena within perturbative expansions have motivated the development of novel mathematical concepts. Since the second half of the 20$^{th}$ century, topological methods have played an increasingly important role in different branches of physics owing to their utility in analysis of high complexity field equations that do not allow a simple general solution. The quantum Hall effect\cite{1981PhRvB..23.5632L}, Aharanov-Bohm's effect\cite{1959PhRv..115..485A}, and the Josephson junction\cite{1974Sci...184..527J} are some examples in which topological arguments help to elucidate the physical origin of the observed phenomena. 

A certain class of non-linear equations allow particle-like solutions, solitary waves or solitons. Thus, a soliton, a freely moving self-trapped state of a non-linear system, can be thought of as a clot of energy concentrated in a small area which is able to move while preserving its shape. Contrary to `true' solitons which preserve their shape during and after collision, there exists so called topological solitons (soliton in a broader sense) which are characterized by a non-trivial topological entity -- topological charge.
These particular solitons are known as topological excitations since their stability purely relies on topological arguments and can not be adiabatically connected to their ground states. The topological excitations are basically determined by the dimension of the space and the order parameter. Thus, for example, in a two-dimensional space, a Heisenberg ferromagnet allows an excitation called the Belavin-Polyakov (BP) monopole\cite{1975JETPL..22..245B}, shown in Fig.~\ref{fig1}a. For this particular example, the BP excitation is equivalent to a skyrmionic particle-like solution\cite{1961RSPSA.262..237S} because of the conformal invariance of the Heisenberg Hamiltonian in two dimensions. In what follows we focus our attention on a skyrmion, a topological soliton corresponding to a non-linear $\sigma$-model.

\begin{figure*}
\includegraphics[width=17.8cm,angle=0]{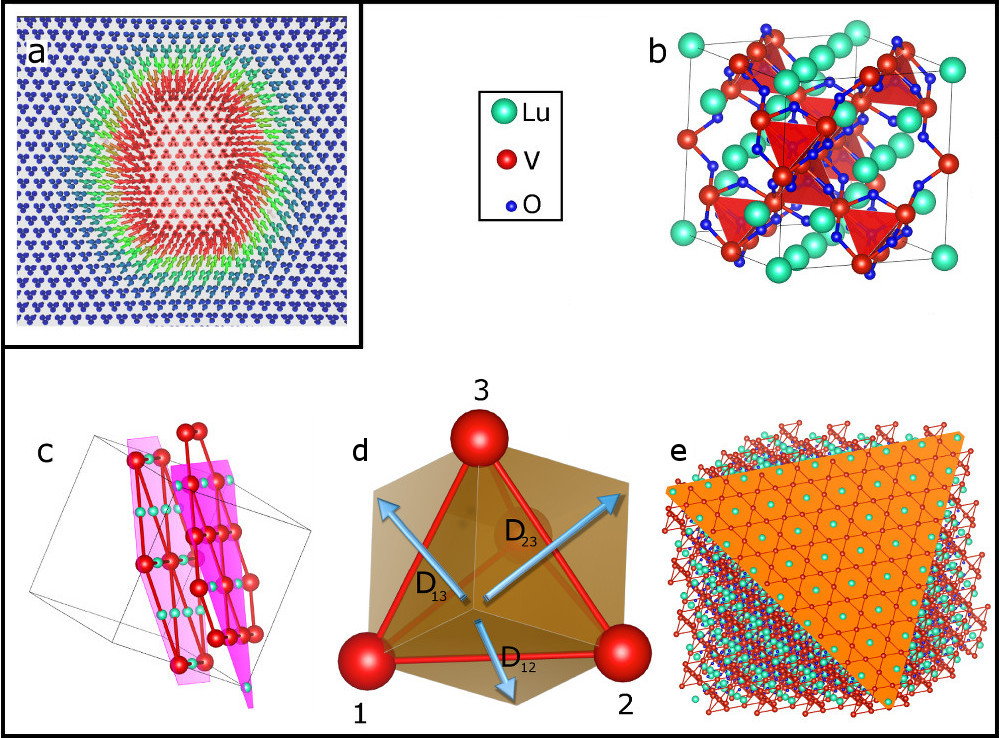}
\caption{\label{fig1}  {\bf Topological magnetic excitations in a Kagom\'e lattice and the crystal structure of Lu$_2$V$_2$O$_7$}. {\bf a} Magnetic texture representing a Belavin-Polyakov monopole or skyrmion. The blue colour represents the atomic spins pointing in the out-of-plane direction while red color means that the magnetization is reversed 180 degrees. The green color represents atomic spins lying in-plane.{\bf b} Pyrochlore structure of Lu$_2$V$_2$O$_7$. {\bf c} Graphical representation of the alternating stacking of triangular and Kagom\'e layers of the pyrochlore lattice.  {\bf d} Schematic representation of the Dzyaloshinskii-Moriya vector for the atoms which form the plane of the Kagom\'e lattice. {\bf e} Vanadium Kagom\'e lattice as a result of cutting a surface-plane perpendicular to the direction [111]. Notice that the color coding of the different atoms is the same in   {\bf b} to {\bf e}.}
\end{figure*}

The concept of the skyrmion originates from the seminal work of T. Skyrme\cite{1961RSPSA.262..237S} who argued that a topologically protected field configuration in a continuous field has particle-like solutions (i.e., skyrmions). A pioneering use of the Skyrme model was employed to describe stable elongated particles (baryons) within the framework of non-linear theory of meson fields. The model is characterized by a conserved baryon number, a topological charge, which is independent of the equations of motion and allows to develop a theory of nucleons obeying a mesonic Fermi-statistic (Bose-fields)\cite{manton}. Within the theory in question, the baryon is nothing but a chiral soliton resulting in collective excitations of pion fields. Its appearance can be attributed to spontaneous breaking of chiral symmetry, while the model itself reveals the nature of Fermi -- Bose transmutation. Thanks to that, the skyrmion became known as a particle-like solution in purely bosonic theories obeying Fermi statistics. Developing this idea introduced a new class of particles with fractional statistics, anyons, which are now widely used. Skyrme's proposal has been extended beyond the scope of high energy physics and in materials science has been an important concept in new and exotic physics, e.g. in nematic liquid crystals\cite{2011NatCo...2E.246F}, ferromagnetic Bose-Einstein condensates\cite{AlKhawaja:2001iq}, high Tc sucperconductivity\cite{2011arXiv1108.3562B}, and quantum Hall magnets\cite{1995PhRvL..75.2562B}. 

Studying low-dimensional magnetic structures remains one of the most challenging and fascinating fields of modern condensed matter physics. Depending on the distance between neighboring spins, crystalline symmetry, and hybridization with substrate, a wide range of magnetic configurations ranging from collinear ferromagnetic, antiferromagnetic and non-collinear helimagnetic configurations to more complicated textures can be observed. If in addition, the inversion symmetry of the system is broken, the spins alignment/configuration gains a certain chirality (handedness) due to the spin-orbit driven antisymmetric exchange interaction, i.e. Dzyaloshinskii-Moriya (DM) interaction. Magnetic skyrmions are chiral spin structures with a whirling configuration so that the plane on which the spins are specified is topologically equivalent to a sphere via, for example, a stereographic mapping. Because of that, a certain topological invariant, namely degree of mapping can be ascribed to the structure. It can be thought of as a skyrmion number as well, providing information analogous to the charge in a particle-like description. One can also evaluate the accumulated Berry phase that influences, {\it e.g.}, traveling electrons when they pass through a skyrmion\cite{2010Natur.465..880P}. 
Recently, it has been shown that the Berry phase gives rise to the DM interaction in the case of smooth magnetic textures when both inversion symmetry is broken and spin-orbit (SO) interaction is present\cite{2013arXiv1307.8085F}. 

A Kagom\'e ferromagnet characterized by a structure with lack of inversion symmetry gives rise to the appearance of DM interaction. In the presence of significantly large DM interactions, the magnon dispersion curves of a Kagom\'e magnet present similarities with the energy band spectra of topological insulators. In fact, in such a system the magnon dispersion relation is gapped in the bulk but allows traveling gapless edge states which are topologically protected against any variations of the material parameters unless the band gap in the bulk collapses. Even though our studied sample is one atomic-layer thick system we denote hereafter bulk states by the states that exist well inside the sample and consequently, by abuse of language bulk skyrmions refers also to the skyrmions that exist inside the 2D sample.

We argue in this communication that this intrinsic property makes a Kagom\'e ferromagnet ideal for creating and controlling the movement of skyrmions. From a technological perspective, skyrmions are promising objects due to their stability. This comes about from the topological nature of the skyrmion, which prevents a continuous deformation into another magnetic configuration with a different topological invariant. Indeed, the existence of topological invariance originates in the duality between $\hat{\bf{k}}$ -- and $\hat{\bf{r}}$ -- space and, as a consequence, when dealing 
with the group velocity, its standard quasiclassical expression gains an anomalous term, which is proportional to the 
so called Berry curvature. Mathematically, the integral of the Berry curvature over a two-dimensional manifold defines an integer invariant, a first Chern class\cite{1999geometry}. For a system which is time-reversal invariant and spatial-inversion invariant, the Berry curvature vanishes. Thus, the Chern number is zero unless the symmetry is broken. In two-dimensional systems, the bands with non-zero Chern numbers must be separated from the other bands by a band gap, such that the boundary between topological phases with different Chern numbers (or other topological invariants) must support the edge modes (similar to the electronic structure at the surface of a topological insulator). Because of the aforementioned properties, the skyrmions are appealing for applications in information storage or logic devices. However, they need to meet several requirements in order to be useful for magnetic applications, namely, they need to have high mobility, small size, and allow for full control of the direction of movement. In this article, we report on the conditions under which skyrmions are created in a Kagom\'e lattice, and we show that this system has all the aforementioned technologically relevant properties. 

The vanadate pyrochlores with generic formula A$_2$V$_2$O$_7$ (A=Lu, Yb, Tm, Y) form a class of ferromagnetic insulators, where the V atoms carry a magnetic moment of about 1.0 $\mu_B$/atom. A very important member of this family is Lu$_2$V$_2$O$_7$, in which very recently the magnon Hall effect has been observed\cite{2010Sci...329..297O} and in addition it was predicted to be a topological magnon insulator\cite{Zhang:2013ws}. In this compound only the V atoms carry a magnetic moment, since Lu is trivalent with a filled 4f shell. The structure of this vanadate (the unit cell is displayed in Fig.~\ref{fig1}b) can be represented as a stacking of alternating Kagom\'e and triangular lattices along the [111] direction. In Fig.~\ref{fig1}c,  we show the stacking of these layers, in such a way that inversion symmetry is broken. As a consequence, the DM interaction is non-zero. According to Moriya's rules\cite{1960PhRv..120...91M} on a single tetrahedron, the DM vector on each bond is parallel to the surface of the surrounding cube and perpendicular to the bond as indicated in Fig.~\ref{fig1}d with blue arrows. We concentrate here our efforts on a model system given by a ferromagnetic Kagom\'e lattice as shown in  Fig.~\ref{fig1}e for the vanadium atoms of Lu$_2$V$_2$O$_7$. The relevant factor in our analysis of topological excitations in the Kagom\'e magnet is represented by the ratio between the antisymmetric ($\mathcal{D}$) and isotropic ($\mathcal{J}$) exchange interactions, namely $\mathcal{D}/\mathcal{J}$. This ratio captures the basic features about the interactions present in the Lu$_2$V$_2$O$_7$ pyrochloride. In experiments\cite{2010Sci...329..297O}, $\mathcal{D}/\mathcal{J}$ has been found to be close to 0.32. Our simulations, described in the methods section, show that the creation and manipulation of skyrmions in the Kagom\'e lattice do not rely too much on small variations of this ratio. Hence, although our results, shown below, are for $\mathcal{D}/\mathcal{J}$=0.4, our theoretical predictions are expected to be relevant for the V atoms forming a Kagom\'e lattice in the pyrochloride Lu$_2$V$_2$O$_7$ as well as systems with similar $\mathcal{D}/\mathcal{J}$-ratio. 

\begin{figure*}
\includegraphics[width=17.8cm,angle=0]{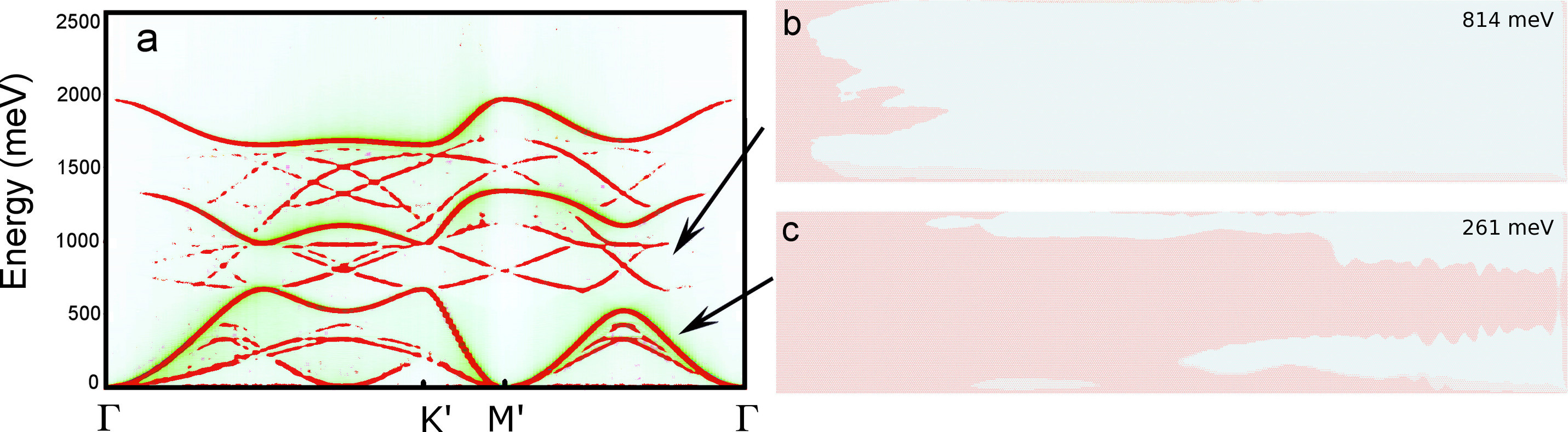}
\caption{\label{fig2}  {\bf Topological magnonic bulk and edge states in a Kagom\'e lattice with $ \mathcal{D}/\mathcal{J}=0.4$}. {\bf a} The dynamic structure factor is shown along the high symmetry points  $\Gamma\rightarrow K^\prime\rightarrow M^\prime\rightarrow\Gamma$, where $K^\prime=2K$ and $M^\prime=2M$. $K^\prime$ and $M^\prime$ are referred to points in the extended Brillouin zone. {\bf b-c} Berry curvature shown in real spin-space averaged over a time interval of 100 ps (see Supplementary Section S2). In {\bf b} and {\bf c}, the system is perturbed with an energy of 814 and 261 meV, respectively. The red colour indicates an excitation where the atomic magnetic moment deviates from the ground state orientation which is represented by the moments pointing along the z axis.  Blue-light colour represent atomic moments oriented along the z axis in a ground state configuration. 
}
\end{figure*}

\section{Results and Discussion}
{\bf Topological bulk and edge states}
We consider here magnetic excitations of a two-dimensional Kagom\'e lattice, e.g. as given by a [111]-surface of the pyrochlore lattice (Fig.~\ref{fig1}e). We start by analyzing bulk and edge states of this system, and to this end we show in Fig.~\ref{fig2}a the magnon dispersion, as revealed from the dynamical structure factor. Note that the results in  Fig.~\ref{fig2}a are obtained using open boundary conditions for a large sample composed of 50$\times$400 unit cells. The intense coloured curves represent bulk magnons, which are seen to be grouped in three branches, with noticeable gaps in-between. Most noteworthy is that in-between the gaps one can find four twisted edge states which form a continuous state. These states cannot be perturbed, so that a gap opens, because they represent edge modes\cite{Zhang:2013ws} that are topologically protected. In fact, the presence of a Dzyaloshinskii-Moriya interaction with a fixed handedness acts as an effective magnetic field in the system. This makes the magnon spectrum of the Kagom\'e lattice looks like the energy band structure of topological insulators. In order to analyze the edge and bulk modes in real space, we show the real-space Berry curvature, averaged over a time-interval of 100 ps, in Figs.~\ref{fig2}b-c  (see Supplementary Section S2 for further details about the calculation of the Berry curvature and topological invariants in the framework of our theoretical method). The results of Figs.~\ref{fig2}b-c were obtained after exciting the first two rows of atoms with an external magnetic field on the left side of the sample, and then the time-evolution of the system was monitored. The excitation was carried out first with an energy of 261 meV and second with an energy of 814 meV. The first excitation represents magnon energies which are allowed in the bulk, whereas the second excitation represents an energy for which bulk states are forbidden, since this energy lies in the magnon-gap. Figure~\ref{fig2}c (261 meV case) displays spin-waves that propagate over the whole sample, corresponding to the expected bulk modes of the 261 meV excitation, whereas Fig.~\ref{fig2}b (814 meV case) shows magnon excitations which only propagate along the edges, i.e. representing edge modes. The data in Fig.~\ref{fig2}a-c demonstrate that magnetic excitations of a Kagom\'e lattice have just as rich physics, in terms of non-trivial topology, as the electronic structure of topological insulators. The advantage with investigations of magnetic excitations is that it is possible to keep track of the real-space information of the magnetic excitation as a function of time, e.g. as given by the information in Figs.~\ref{fig2}b and \ref{fig2}c.

\begin{figure*}
\includegraphics[width=17.8cm,angle=0]{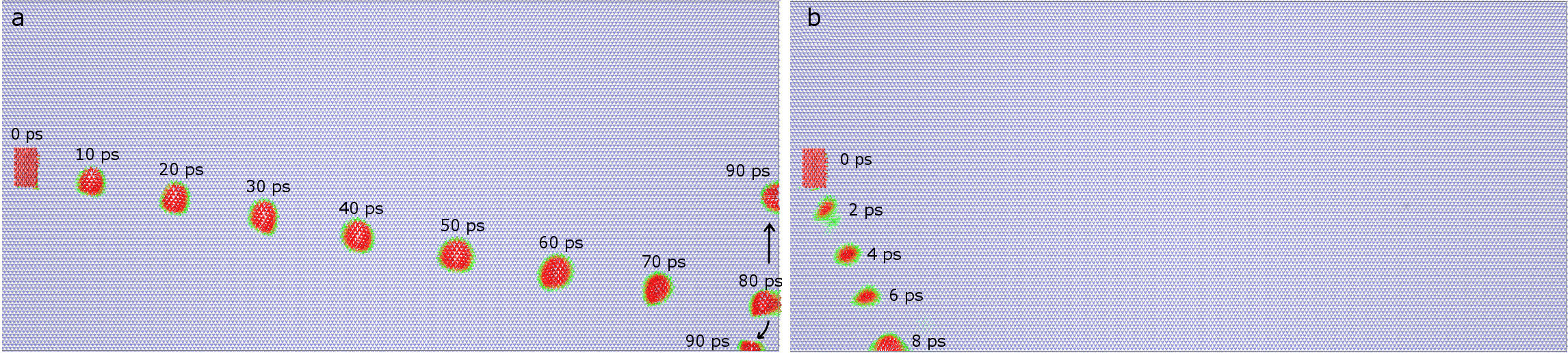}
\centering

\caption{\label{fig3}  {\bf Snapshots of a skyrmionic excitation created in the center of the Kagom\'e stripe, at T= 1mK}. {\bf a} Time evolution of a SA pair, and the generation of a meron and antimeron after colliding with the edge of the sample. The measured SA pair speed is about 1290 m/s. (Supplementary video 1).{\bf b} Same simulation conditions as in {\bf a}, but with different direction of the linear momenta of the SA pair, due to the non-linear process of the SA pair generation (see text and Supplementary video 2). Same color convention as in Fig.~\ref{fig1}a applies in this figure and throughout the rest of this paper. In both figures, we only show the right half side  of the Kagom\'e stripe so that the rectangular excitation at 0 ps is located in the center of the stripe.    
}
\end{figure*}

{\bf Local topological excitations}
Next, we consider local excitations, e.g. as they would emerge from a spin-transfer torque of a spin-polarized STM tip or as described in the experiments of Ref.~\cite{2013Sci...339.1295M}. This excitation was generated by applying a local torque that causes all the atomic spins subjected to this torque to have their moments reversed with respect to the majority of the spins in the simulation cell (Fig.~\ref{fig3} shows the configuration at t=0 ps). We then instantaneously removed this torque and observed that the resulting excited state could be thought of as skyrmion-like or a skyrmion/anti-skyrmion (SA) pair, that in most simulations were stable for long times ($>$ 100 ps). This procedure was repeated several times and it was found that skyrmionic excitations were always stabilized. Figure~\ref{fig3} shows the generation and time evolution of two such skyrmionic excitations (a movie of these excitations is found in Supplementary videos 1 and 2). It may be seen that they are both stable over long times but that the initial direction of their movement is different (see Figs.~\ref{fig3}a-b).

The size and magnitude of the linear momentum of the skyrmionic excitation is in general found to be stochastic and depends on minute details of the conditions of the spin-system just after removing the local torque. As discussed above, this is obvious when comparing Fig.~\ref{fig3}a and Fig.~\ref{fig3}b. In both figures, the simulation parameters were exactly the same, but the highly non-linear process of the creation of the SA pair, combined with small thermal fluctuations, edge effects on the spin-reversed region at the beginning of the simulation etc., causes the direction of the linear momentum of the skyrmionic excitation to be very different. Furthermore, in Fig.~\ref{fig3}b, the excitation is found to experience stronger damping, in a process which emits spin-waves. As a result of the new direction, the SA pair collides with the edge 8 ps after it is created, which is a very short period of time for the SA pair to decrease its linear momentum and consequently, the collision destroys the SA pair and produces spin waves. The results in Fig.~\ref{fig3} demonstrate that SA pairs are easily generated but that there is an uncertainty in the determination of their linear momentum and their life-time.

The competition between exchange and DM interaction favours canted magnetization and as a result, the effects associated with finite size geometry are of crucial importance. A non-trivial magnetic texture is in general subjected to the Magnus force, which prevents it from moving to the boundary. However, it can be easily shown that as long as the strength of a local torque is sufficiently small, a repulsion from the edge is circumvented. 
In Fig.~\ref{fig3}a, a coupled SA pair has time to become stabilized ($\sim$ 80 ps) and is found to travel with a speed of $\sim$ 1290 m/s to the right edge of the sample. The linear momentum is large enough for the SA pair to reach the edge, where it splits into a separate meron and antimeron. Both the meron and antimeron are stable during a significant period of time, i.e. at least longer than 10 ps.
 
The results in Fig.~\ref{fig3} motivate an analysis in terms of a meron\cite{1976PhLB...65..163D} (in our case, half-skyrmion), that originally was introduced to resolve the problem of confinement in particle physics, namely when quarks are bound by strong forces at large distances \cite{Rajaraman:1982tv}. The latter fact prevents the application of a semiclassical approximation directly. In fact, within the semiclassical approach one may look for a perturbative expansion around the minima of the Euclidean action (this effectively leads to a series obtained by re-summing an infinite number of graphs). Thus, using this formalism restricts one to finite-action solutions, also known as instantons. However, elevated temperatures result in higher-energy configurations being involved, possibly with infinite-action. An elegant way to take into account such terms \cite{1978PhRvD..17.2717C} suggests to add so called meron solutions. In fact, originally proposed in the framework of Yang-Mills theory \cite{1976PhLB...65..163D}, the meron is characterized by a non-integer topological charge, namely $1/2$. In this spirit, a finite-action solution with topological charge 1 can be made of two merons with finite separation (a bound meron pair) which breaks up into constituents when large distance effects become prominent. Hereafter, the terms half-skyrmion or meron and half-antiskyrmion or antimeron will be considered as synonymous, respectively.

\begin{figure}
\includegraphics[width=8.5cm,angle=0]{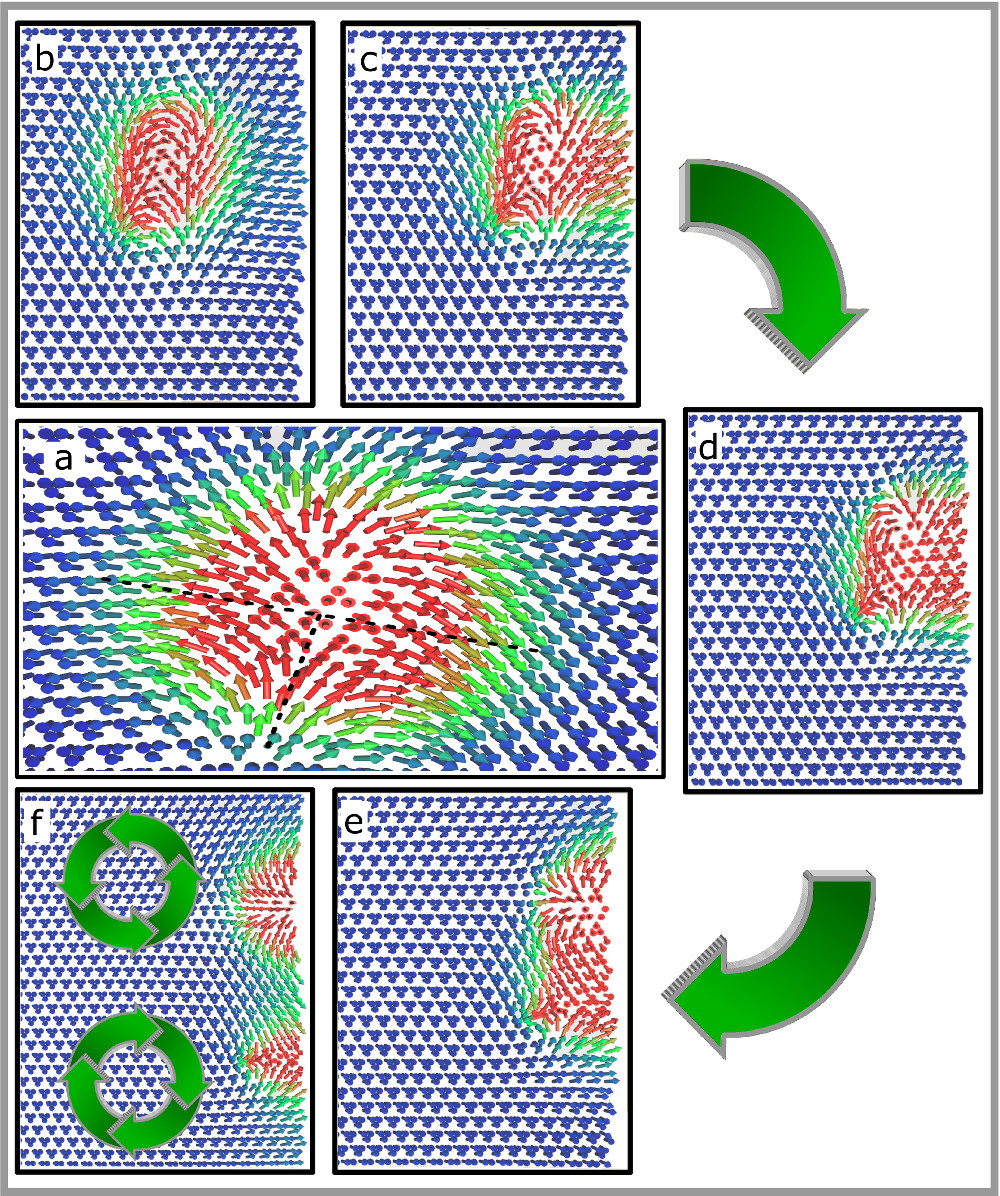}
\caption{\label{fig4}  {\bf Snapshots of a coupled skyrmion-antiskyrmion pair colliding with the edge of the Kagom\'e stripe}. {\bf a} Illustration of a skyrmion-antiskyrmion pair before the collision with the edge of the Kagom\'e stripe. Dashed lines are a guide to the eye for recognizing the antiskyrmion magnetic texture while the other half of the magnetic excitation resembles a magnetic texture like in the Belavin-Polyakov monopole. {\bf b-f} Several frames showing the coupled SA pair colliding with the edge of the stripe and the resulting SA decoupling due to the chiral edge states. The snapshots were taken from the same simulation displayed in Fig.~\ref{fig3}a.     
}
\end{figure}

In general, a skyrmion that moves towards an edge becomes annihilated. However, for the Kagom\'e magnet the annihilation of the skyrmion does not occur once it reaches the edge, since the existence of chiral magnonic edge states gives rise to profound changes in the excitations of the Kagom\'e lattice, as we have seen in Fig.~\ref{fig2}. Thus, once the SA pair reaches the edge, the chirality forces them to be separated as two distinct entities. As shown in Fig.~\ref{fig3}a, the meron and antimeron emanating from the collision of the SA pair at the edge of the Kagom\'e lattice have drastically different speeds, since the distance traveled by the antimeron is shorter than the one of the meron during the same period of time. A magnification of the SA pair discussed in Fig.~\ref{fig3}a is shown in Fig.~\ref{fig4}. Notice that because of the chosen magnetic orientation of the spins studied here, the meron has counter-clockwise chirality contrary to the antimeron which has clockwise chirality. In Fig.~\ref{fig4}a we show the coupled SA pair, just before it reaches the edge. Note that the dashed line shows the magnetic texture that has a half-antiskyrmion. Hence this illustrates how the SA pair is coupled before it reaches the edge. Figures~\ref{fig4}b-f show the time-evolution of this SA pair over a time interval of 90 ps, just before (Fig.~\ref{fig4}b) and just after (Figs.~\ref{fig4}c-f) it reaches the edge. Note that once the coupled SA pair reaches the edge, it becomes unstable, due to non-trivial topology, and breaks up into a separated meron and antimeron (Figs.~\ref{fig4}e and f), that then travel along the edge in a decoupled fashion and in opposite directions.

The results of Figs.~\ref{fig3} and \ref{fig4} demonstrate that it is possible to create SA pairs in a Kagom\'e lattice and that they are stable over substantial times. It is also clear that the SA pairs can travel with supersonic velocities of about 1300 m/s. However, the linear momentum of such SA pairs is unfortunately difficult to control or design by the initial conditions of their generation. A way to overcome this problem is to make use of the topological properties of the edge states of the Kagom\'e magnet and to place the local excitation at the edge of the sample. In Fig.~\ref{fig5} and in Supplementary videos 3 and 4, we illustrate as an example the generation and time evolution of a skyrmion and antiskyrmion pair, created in exactly the same way as described in Fig.~\ref{fig3}. In Fig.~\ref{fig5}a it is shown that the meron moves to the right side whereas the antimeron travels in the opposite direction. The difference in their direction is produced by their different chirality. Local excitations placed at the edge do not always generate SA pairs; it is quite possible that single merons are generated. As an example we show in Figure~\ref{fig5}b a singular meron (the antimeron is indeed created also here, but is damped out very quickly).

The important message from the results of Fig.~\ref{fig5} is that the direction of the meron and antimeron is totally controlled at the edge of the Kagom\'e lattice. Indeed, the topological nature of these excitations prevents the meron or antimeron to leave the edge, or to collapse. The stability of these chiral excitations is illustrated further by the fact that they can survive traveling along a 90 degree corner, as is shown in Fig.~\ref{fig5}a and Supplementary video 3, for an antimeron. The strong stability and long life-times of these excitations become promising when considering applications, because magnetic information can be transported over long distances along complicated paths and for long times. Our results suggest that it is quite possible to create a meron (write) at one point of space and time and to detected it (read) at another distant point along the edge. This property paves the way for novel applications of skyrmions like, for example, performing logical operations (skyrmionics). 

\begin{figure*}
\includegraphics[width=17.8cm,angle=0]{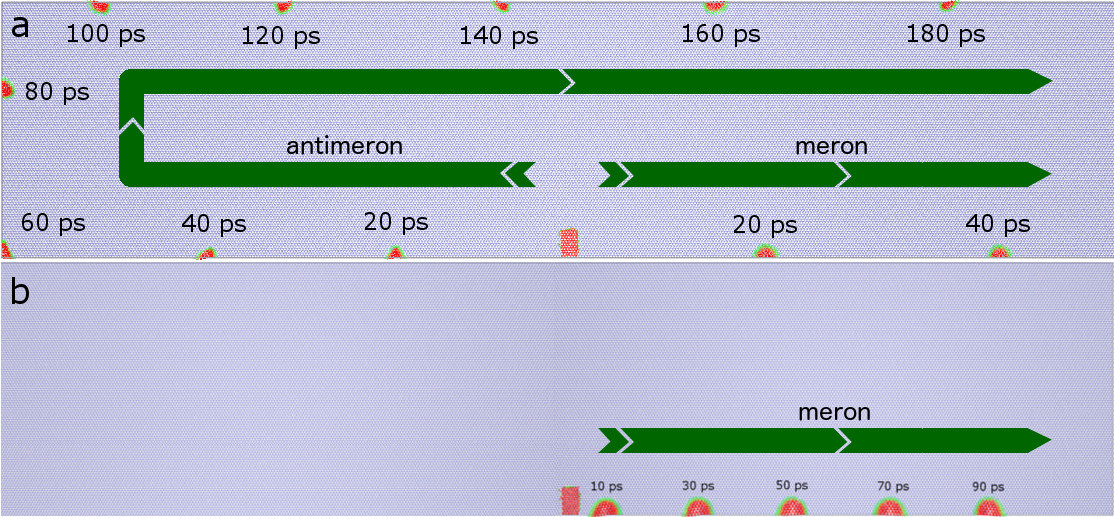}
\caption{\label{fig5}  {\bf Snapshots of skyrmionic excitations at the edge of the Kagom\'e stripe}. {\bf a} Generation and stabilization of a meron and an antimeron. In this case, the antimeron is able to overcome a 90 degree bend. (Supplementary video 3). {\bf b} Generation and time evolution of a single meron. The antimeron is damped out very quickly after it is generated. (Supplementary video 4). In both cases (a and b), the meron moves counterclokwise and the antimeron clockwise.
}
\end{figure*}

\begin{figure*}
\includegraphics[width=17.8cm,angle=0]{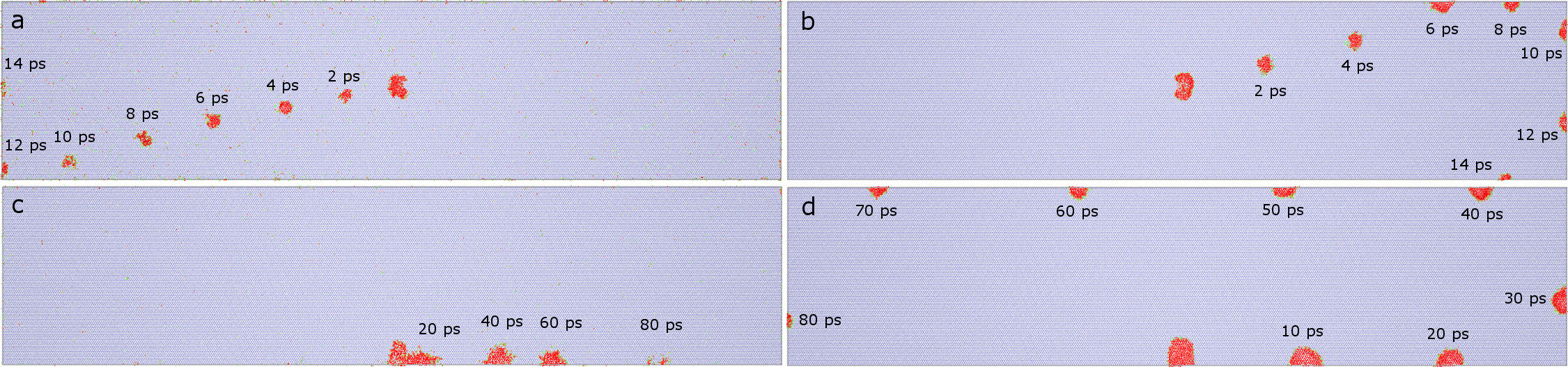}
\caption{\label{fig6}  {\bf Snapshots of skyrmionic excitations at room temperature}. In ({\bf a}) we show the creation of a SA pair and in ({\bf c}) an edge meron, using $\mathcal{J}$=5 mRy and $\mathcal{D}$=2 mRy (Supplementary video 5). In {\bf b} we show a SA pair and in {\bf d} an edge meron, using $\mathcal J$=10 mRy and $\mathcal{D}$=4 mRy (Supplementary video 6). 
}
\end{figure*}

Recent experimental results on chiral bulk magnets like MnSi \cite{2009Sci...323..915M} and Fe$_{0.5}$Co$_{0.5}$Si \cite{2010Natur.465..901Y} have indeed identified a skyrmion lattice phase. These works came to the conclusion that the so-called skyrmionic ``A phase'' is stabilized in a very limited region of a phase diagram defined by the applied magnetic field and temperature of the sample.
The onset of the skyrmionic ``A phase'' is observed only in a narrow window of this phase diagram, basically at B=20 mT and T=25-30 K. Spin polarized scanning tunnelling experiments of Fe monolayers grown on Ir(111) also show that skyrmions are observed only at very low temperatures (T=11 K) \cite{2011NatPh...7..713H}. As far as we know, the highest temperature for the skyrmionic phase has been reported in Ref.~\cite{yu}, where the skyrmions have been manipulated with electrical currents at temperatures up to 275 K in FeGe. All in all, this makes these excitations less suitable for room temperature applications. However, our theoretical considerations outlined above, show that SA pairs generated in a Kagom\'e lattice do not have this limitation. As an example, we show in Fig.~\ref{fig6} and Supplementary videos 5 and 6, the generation and movement of bulk and edge skyrmions at a temperature of 300 K. Figures~\ref{fig6}a and~\ref{fig6}b show the creation and temporal evolution of bulk skyrmions, whereas Figs.~\ref{fig6}c and~\ref{fig6}d show edge skyrmions. Note that the strength of the exchange and DM interaction in Figs.~\ref{fig6}b and~\ref{fig6}d is twice as large as that used in Figs.~\ref{fig6}a and ~\ref{fig6}c and that skyrmionic excitations seem to be insensitive to the actual strength of the exchange and DM interaction, as long as their ratio is close to 0.4. We note here that we also performed simulations for slightly different ratios between the strength of the exchange and DM interaction, with very similar results as shown in Figs.~\ref{fig3}-\ref{fig6}.

\begin{figure}
\includegraphics[width=8.5cm,angle=0]{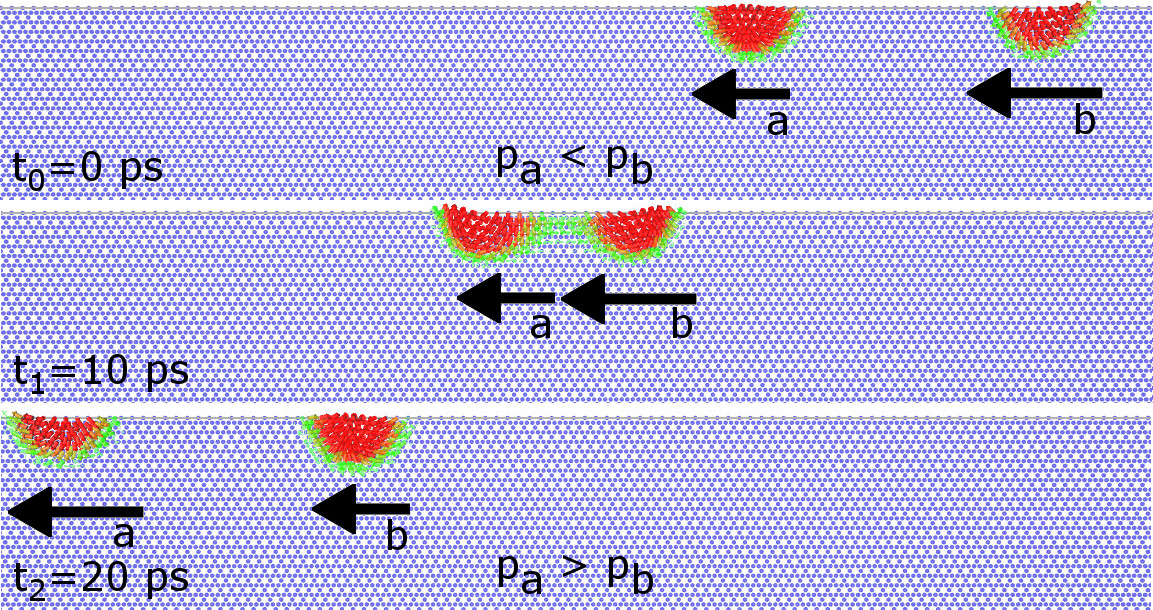}
\caption{\label{fig7}  {\bf Snapshots of the meron-meron collision}. At $t_0$, meron {\bf b} has higher linear momentum (${\bf p}_b$) than meron {\bf a} (${\bf p}_a$) but 10 ps after the collision ($t_2$), meron {\bf a} moves away from meron {\bf b} with higher velocity. (Supplementary video 7)
}
\end{figure}

{\bf Collision of topological excitations}
Our results so far point out that it is possible to create skyrmionic excitations that can move with significant speed along edges and it is natural to ask if collisions between such particle-like excitations can happen, and if so, how do they happen? To unravel this question, we have studied meron-meron and meron-antimeron collisions (at T= 1mK). This was done by creating, simultaneously, two excitations at the edge of the Kagom\'e lattice. The excitations were created with sufficient distance, in order to avoid any initial interaction or correlation between them due to the emission of short lifetime spin waves generated during the excitation of the system. Once the two edge skyrmions were excited they started to travel counter-clockwise along the edge of the lattice, the first (meron {\bf a}) being ahead of the second (meron {\bf b}). For this particular pair of merons, the second one had higher speed, hence allowing for investigating a meron-meron collision. In Fig.~\ref{fig7}, three snapshots show the moment before ($t_0$), during ($t_1$) and after ($t_2$) this collision.   As the figure and Supplementary video 7 shows, the two merons seems to experience an elastic collision, so that before the collision ($t_0$) meron {\bf b} has a linear momentum that is larger than that of meron {\bf a}. At the collision ($t_1$) the meron {\bf b} provides linear momentum to meron {\bf a}, in such a way that 10 ps after the collision ($t_2$), meron {\bf a} has a higher momentum than meron {\bf b}. 

The results of Fig.~\ref{fig7} pose questions regarding the details of a seemingly elastic collision. We analyze this by considering that the dynamics of the meron traveling along the x-direction can be described by the Lagrangian $\mathcal{L}=\frac{m}{2}(\partial_t x)^2$, 
where x is the coordinate of the meron center of mass and m is the ``mass'' of the meron. The ``mass'' is considered to be equal to the number of spins constituting the meron and the momentum is then $P_x=\frac{\partial\mathcal{L}}{\partial(\partial_t x)}=m\partial_t x$. It should be noted that it is not possible to exactly specify the boundary of a meron due to the discrete nature of the spin-lattice. Hence the calculation of the ``mass'' of a meron is associated with a small numerical error. Nevertheless, the calculated masses for the merons in Fig.~\ref{fig7}, before the collision, are m$_a\simeq 176$ spins and m$_b\simeq 210$ spins, while after the collision, they change the mass slightly, becoming m$_a\simeq 220$ spins and m$_b\simeq 182$ spins. We have also evaluated the velocities of the merons just before and after the collision and hence been able to analyze the linear momentum of the collision in Fig.~\ref{fig7}. The total linear momentum before the collision is about $P_x\simeq 4652$ (\AA~spins)/ps and after the collision it is $P_x\simeq 4620$ (\AA~spins)/ps. Moreover, we have also computed the kinetic energy before and after the collision and found that it is 11.40 and 11.30 (\AA/ps)$^2$spin, respectively. Hence, within numerical errors, both the linear momentum and kinetic energy are conserved, demonstrating the elastic nature of the collision in Fig.~\ref{fig7}. In contrast to elastic collisions of purely classical systems the mass of an individual meron is not conserved during the collision in Fig.~\ref{fig7}. 

\begin{figure*}
\includegraphics[width=17.8cm,angle=0]{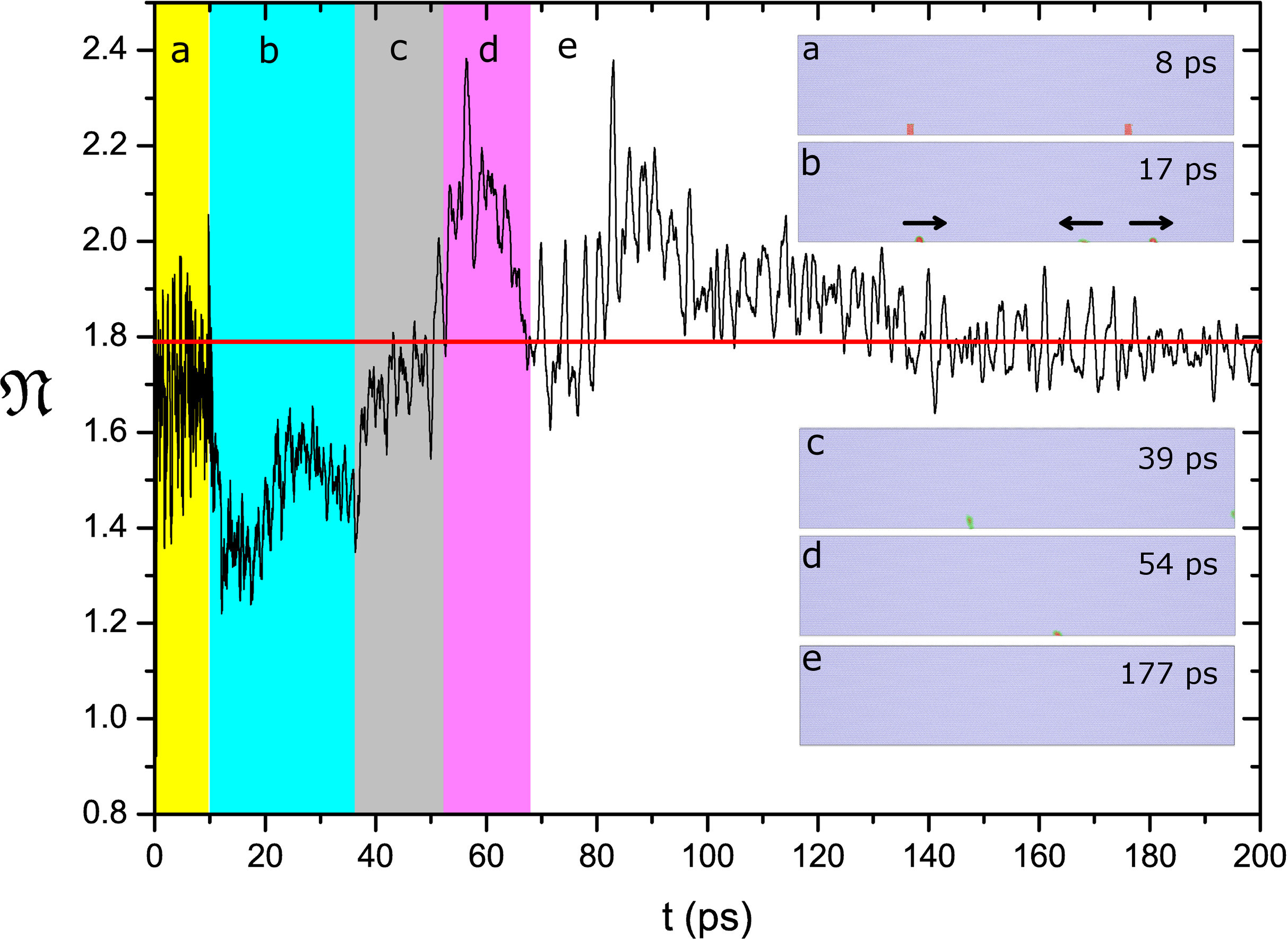}
\caption{\label{fig8}  {\bf Snapshots at different stages of the meron-antimeron annihilation process}. {\bf a} Two regions at the lower edge of the Kagom\'e stripe with reversed magnetisation, due to a local applied torque. {\bf b} Generation of two merons and one antimeron after removing the local torque from the surface. {\bf c} Meron-antimeron annihilation collision. {\bf d} Generation of a second meron as a result of the emission of spin-waves in the annihilation collision in {\bf c}. {\bf e} Equilibration of the spin-waves. The figure also illustrates the time dependence of the total skyrmion number $\mathfrak{N}$.  The horizontal red line indicates the skyrmion number at the first time step (Supplementary video 8).
}
\end{figure*}

It is possible also to investigate meron-antimeron collisions and we show in Fig.~\ref{fig8} several snapshots of the generation of a meron and an antimeron and their time evolution. Figure~\ref{fig8} and Supplementary video 8 shows that once created, they move towards each other. Figure~\ref{fig8} also shows the time-evolution of the skyrmion number of the whole simulation cell. The skyrmion number was calculated as shown in the Methods section at every simulation time step. Similar to the generation of the separated pair of skyrmionic excitations (Fig.~\ref{fig7}), the skyrmion - anti-skyrmion pair was generated at the edge, by first reversing moments in a region by a local torque and then (after 10 ps) removing this torque rapidly. Looking at the details of Fig.~\ref{fig8} we note that initially a meron-antimeron is generated at the right hand side of the sample and only a single meron is generated on the left side. We hence have three dynamical excitations which we will refer to as the right meron (from the right side of the sample), the left meron (from the left side of the sample) and the antimeron. Figure~\ref{fig8} shows that the right meron gets damped after a while and leaves the system. The left meron and the antimeron are more long-lived and are found to move towards each other and collide after 39 ps (Fig.~\ref{fig8}c). At this collision, they fuse and destroy each other. As a result, the energy is carried out by outgoing spin waves, which after 54 ps create a meron in a totally different part of the lattice (Fig.~\ref{fig8}d) that subsequently moves to the right along the edge before it gets damped and leaves the system. This process hence shows a very non-localized process of the destruction and creation of skyrmionic excitations. 

Figure~\ref{fig8} also shows that the skyrmion number ($\mathfrak{N}$) is not an integer and it is not very smooth as a function of time. This behaviour is due to the fact that the skyrmion number was calculated for the whole sample and we used a very low damping. In consequence, there are contributions to $\mathfrak{N}$ arising also from regions far away from the place where the skyrmionic excitations are located. We can nevertheless get some insight from this number if we concentrate on the difference of this parameter at certain time steps. 
The horizontal red line indicates the value of the skyrmion number for the initial step, so we take this value as the background skyrmion number. During the first 10 ps (yellow region), the applied local torques are present.  
In region b, just after the creation of the topological excitations, the skyrmion number is reduced with roughly 0.5. This is because we have created an antimeron with skyrmion number -0.5. In the transition b$\rightarrow$c, the meron-antimeron collision is produced and they annihilate each other while the remaining meron shown on the right side dies after colliding with the vertical edge. In region c, the skyrmion number tends to come back towards the ground value with a jump of about 0.5 because the antimeron was annihilated. In this region we still can appreciate the remaining meron just before dying. In region d, the increase of the skyrmion number of about 0.5 is due to the creation of another meron. This excitation is created by spin-waves that were emitted during the meron-antimeron annihilation. This object is however damped away very quickly with strong spin-wave activity during the first part of region e. After some time, the system equilibrates and the skyrmion number goes back to the initial value as it is observed in Fig.~\ref{fig8}e.

{\bf Choice of materials and concluding remarks}
As regards suitable materials for which the here predicted phenomena should be possible to observe at room temperature, we note that larger values of $\mathcal{J}$ and $\mathcal{D}$ have a better chance for stabilizing skyrmions. However, the skyrmionic excitations discussed here are stable for a wide range of $\mathcal{J}$ and $\mathcal{D}$. The exchange interaction considered here ranged from 1 up to 10 mRy in steps of 1 mRy while $\mathcal{D}$ was varied from 0.4 till 4 mRy in steps of 0.4 mRy. Hence, we choose to keep $\mathcal{J}/\mathcal{D}=0.4$, since this ratio is known from previous experimental investigations. Skyrmionic phases show up when $\mathcal{J}\geq5$ mRy and $\mathcal{D}\geq2$ mRy (e.g. as shown in Fig.~\ref{fig6}). For cases in which $\mathcal{J}<5$ mRy, skyrmions are still stable but their shape is not well defined and skyrmionic excitations with lower values of $\mathcal{J}$ and $\mathcal{D}$ have higher probability to be damped away just after their creation. As shown in Figs.~\ref{fig6}b and d, larger values of $\mathcal{J}$ and $\mathcal{D}$ results in skyrmions that are more stable and less susceptible from thermal fluctuations. 

We have outlined conditions for novel exotic magnetic excitations of the Kagom\'e lattice. The possibility to observe bulk and edge states is outlined here, where in particular, the edge states have non-trivial topological properties that provide unique magnon modes. We have also shown that the Kagom\'e lattice has potential to exhibit SA pairs, meron and antimeron excitations even at room temperature. This is found for a wide range of parameters describing the magnetic excitations, in particular, the Dzyaloshinskii-Moriya interaction and the Heisenberg exchange interaction, as long as the ratio between them is in the range 0.3-0.4. Such a value of the ratio between these two magnetic interactions, is close to the value found in experiments\cite{2010Sci...329..297O}.
  
Due to the non-trivial topology of the edge-states, we find that it is possible to control the movement of skyrmionic excitations and that such excitations are long lived. We have even found that these particle like-excitations can move on well defined straight lines and even be made to turn around corners. This offers a potential to use such excitations in different emerging technologies like for example in data storage or manipulation of magnetic bits. We have also investigated  meron-meron collisions as well as meron-antimeron collisions. The former are found to undergo basically elastic collisions, whereas the latter are more complex, where annihilation of a meron-antimeron collision is followed by the birth of a new meron, a process which occurs as a highly non-local phenomenon. 
 
Pyrochlores are possible candidates for observing the phenomena predicted in this work and particularly we discussed here the 111-cut of the vanadate pyrochlores. The Kagom\'e lattice can however be found in many other compounds, e.g. SrCr$_8$Ga$_4$O$_{19}$\cite{obradors}, Ba$_2$Sn$_2$Ga$_3$ZnCr$_7$O$_{22}$\cite{cava} and the jarosites like KM$_3$(OH)$_6$(SO$_4$)$_2$\cite{keren,lee,inami} with M = V, Cr, or Fe. These materials are also potential candidates for investigations of non-trivial topological magnon edge-states and skyrmionic excitations.

\section{Methods}

Our theoretical methodology\cite{2008JPCM...20E5203S} consists of mapping an itinerant electron system onto an effective spin-Hamiltonian\cite{2003PhRvB..68j4436U,1987JMMM...67...65L},
\begin{equation}
\label{hamiltonian}
\mathcal{H}=\sum_{<ij>}\left[-\mathcal{J}_{ij} {\bf s}_i \cdot{\bf s}_j+ \boldsymbol{\mathcal{D}}_{ij} {\bf s}_i \times {\bf s}_j\right]-g\mu_B {\bf B} \sum_i {\bf s}_i
\end{equation}
where $<$$ij$$>$ denotes the atomic indices to the first neighbour pairs, $s_i$ is the atomic moment, $\mathcal{J}_{ij}$ is the strength of the exchange interaction and the spin-orbit contribution is included via the strength of the DM vector $\boldsymbol{\mathcal{D}}_{ij}$. The last term comes from the Zeeman effect under an external magnetic field $\bf{B}$, where g is the g-factor and $\mu_B$ represents the Bohr magneton. We consider that the exchange-interaction as well as the modulus of the DM vector are the same for every pair of atoms, which is consistent with all atoms of the Kagom\'e lattice being of the same type. Hence, we use $\mathcal{J}_{ij}=\mathcal{J}$ and $|\boldsymbol{\mathcal{D}}_{ij}|=\mathcal{D}$.

In order to capture the dynamical properties of spin systems at finite temperatures we used an atomistic spin dynamics (ASD) approach\cite{2008JPCM...20E5203S}. The equation of motion of the classical atomistic spins at finite temperature is governed by Langevin dynamics via a stochastic differential equation, normally referred to as the atomistic Landau-Lifshitz-Gilbert (LLG) equations, which can be written in the form; 
\begin{eqnarray}
\frac{\partial s_i}{\partial t}&=&-\frac{\gamma}{1+\alpha_i^2} {\bf s}_i \times [{\bf B}_i+{\bf b}_i(t)]\nonumber\\&-&\frac{\gamma \alpha_i}{s (1+\alpha_i^2)} {\bf s}_i \times {\bf s}_i \times [{\bf B}_i+{\bf b}_i(t)] 
\end{eqnarray}
where $\gamma$ is the gyromagnetic ratio and $\alpha_i$ denotes a dimensionless site-resolved damping parameter which accounts for the energy dissipation that eventually brings the system into a thermal equilibrium. The effective field in this equation, is calculated as ${\bf B}_i=-\partial \mathcal{H}/\partial {\bf s}_i$ and the temperature fluctuations (T) are considered through a random Gaussian shaped field ${\bf b}_i(t)$ with the following stochastic properties:
\begin{eqnarray}
\langle b_{i,\mu}(t)\rangle&=&0, \nonumber \\
\langle b_{i,\mu}(t) b_{j,\nu}(t')\rangle&=&\frac{2\alpha_i k_B T \delta_{ij}\delta_{\mu\nu} \delta(t-t')}{s(1+\alpha_i)^2\gamma } 
\end{eqnarray}
where $i$ and $j$ are atomic sites, while $\mu$ and $\nu$ represent cartesian coordinates of the stochastic field. After solving the LLG equations, we have direct access to the dynamics of the atomic magnetic moments $s_i(t)$ and with this information we can calculate all relevant dynamical properties like the connected space- and time-displaced correlation function defined as
\begin{equation}
C^k({\bf r}-{\bf r}^\prime,t, t^\prime)=\langle s_r^k(t)s_{r^\prime}^k(t^\prime)\rangle-\langle s_r^k(t)\rangle \langle s_{r^\prime}^k(t^\prime)\rangle
\end{equation}
where $\langle\cdots\rangle$ denotes an ensemble average and k is the cartesian component. The magnon dispersion relation can be evaluated via the dynamic structure factor $S({\bf q},\omega)$, which is nothing else than the space and time Fourier transform of the connected correlation function,
\begin{equation}
S^k({\bf q},\omega)=\frac{1}{N}\sum_{\bf{r},\bf{r}^\prime} e^{-i{\bf q}\cdot({\bf r}-{\bf r}^\prime)}\int_{-\infty}^{\infty}e^{i\omega t} C^k({\bf r}-{\bf r}^\prime,t)\;\mathrm{d}t
\end{equation}
where $N$ is the number of terms in the summation while ${\bf q}$ and $\omega$ are the momentum and energy transfer (see Supplementary Section S1 for further details about the simulations).

The skyrmion number $\mathfrak{N}$ represents a topological index of the field configuration and is defined and evaluated here, by
\begin{equation}
\label{sknumber}
\mathfrak{N}=\int_\mathbb{S} {\bf s}\cdot \left(\frac{\partial {\bf s}}{\partial x}\times\frac{\partial {\bf s}}{\partial y}\right)\; \mathrm{d}x \mathrm{d}y
\end{equation} 
where $\mathbb{S}$ represents a two-dimensional compact orientable manifold. The topological excitation with $\mathfrak{N} > 0$ is called a skyrmion while in the case of $\mathfrak{N} < 0$ it is termed an antiskyrmion. The same formalism can be extended straightforwardly to describe multiskyrmionic states and in particular skyrmion-antiskyrmion pairs\cite{Komineas:2007ur}.

\section{Acknowledgements}
The authors thank the European Research Council (ERC Project No. 247062-ASD), the Swedish Research Council (VR), the Knut and Alice Wallenberg Foundation for financial support. A.B. and O.E. acknowledge support from eSSENCE. The computer simulations were performed on resources provided by the Swedish National Infrastructure for Computing (SNIC) at the National Supercomputer Centre (NSC), Chalmers Centre for Computational Science and Engineering (C3SE) and High Performance Computing Center North (HPC2N).

\section{Supporting Information: \\Topological excitations in a Kagom\'{e} magnet}

\vspace{0.5cm}
\noindent {\bf S1: Atomistic spin model}

\vspace{0.5cm}

The atomistic modelling used in this work has recently been developed in our group. It basically follows standard techniques in this research area. We include a brief description of the model here for completeness but more details can be found in Ref.~{\cite{2008JPCM...20E5203S}}. The method relies on solving the Landau-Lifshitz-Gilbert stochastic differential equation of motion
\begin{equation}
\label{LLG}
\frac{\partial s_i}{\partial t}=-\frac{\gamma}{1+\alpha_i^2} {\bf s}_i \times [{\bf B}_i+{\bf b}_i(t)]-\frac{\gamma \alpha_i}{s (1+\alpha_i^2)} {\bf s}_i \times {\bf s}_i \times [{\bf B}_i+{\bf b}_i(t)] 
\end{equation}  
where $\gamma$ is the gyromagnetic ratio and $\alpha$ is the dimensionless site-resolved damping parameter which accounts for the energy dissipation and eventually brings the system into a thermal equilibrium.  The effective field ${\bf B}_i$ acting on every spin ${\bf s}_i$ is calculated as ${\bf B}_i=-\partial \mathcal{H}/\partial {\bf s}_i$, where $\mathcal{H}$ represents a parametrized Hamiltonian which can accounts for several terms including both interatomic isotropic (Heisenberg) and anisotropic (Dzyaloshinskii-Moriya) exchange interactions, magnetocrystalline anisotropy, dipolar interaction, Zeeman term and many others. The coupling of the spin system with a thermal reservoir is through Langevin dynamics so that the spin excitations are generated by performing classical rotations in such a way that the spin energy satisfy Boltzmann statistics. Thus, the thermal fluctuations are represented here by $b_i(t)$ which is nothing else than a random Gaussian shaped field with the following stochastic properties
\begin{eqnarray}
\langle b_{i,\mu}(t)\rangle&=&0, \nonumber \\
\langle b_{i,\mu}(t) b_{j,\nu}(t')\rangle&=&\frac{2\alpha_i k_B T \delta_{ij}\delta_{\mu\nu} \delta(t-t')}{s(1+\alpha_i)^2\gamma } 
\end{eqnarray}
where $i$ and $j$ are atomic sites while $\mu$ and $\nu$ represent cartesian coordinates. The equation of motion is integrated with the semi-implicit midpoint solver using a time step of 0.1 fs to ensure numerical stability\cite{2010JPCM...22q6001M}. The flowchart of the UppASD code \cite{2008JPCM...20E5203S} follows basically three steps. In the initialization step, all the parameters necessary to describe the system (geometry, dimensions, boundary conditions, ...) are set up. During the optional second step, also called initial phase, the system is brought into thermal equilibrium while in the last step (measurement phase) the system is evolved in time with a complete data sampling being made. For bringing the system into thermal equilibrium (thermalization), we have performed a Monte Carlo simulation with Metropolis dynamics\cite{2009gmcs.book.....L}. We used five simulated annealing steps in order to bring the system to the agreed temperature. After that, spin dynamics simulations solving Eq.~(\ref{LLG}) has been performed to properly describe the spin dynamics of the system. The material parameters along with the simulation parameters we use in the model are given in Tables~\ref{table1} and~\ref{table2} for T=0.001 K and T=300 K, respectively. We used a large sample (50$\times$400 unit cells) to ensure that the statistical calculations as well as the numerical stability of Fourier transform in this system is reliable. According to the size of the sample, the total number of spins used in the simulations is 127200. In Fig.~\ref{fig1} we show the unit cell we have used for setting up the Kagom\'e lattice. Thus, we ensure that our Kagom\'e stripe has four 90 degree corners. This point is pretty much important because it allow us to asses the stability of the skyrmions when they collide with a 90 degree corner. In most of the cases, they were able to overcame this difficulty and consequently, it proves that the topology protect them against structural deformations. Moreover, we also set up the system with open boundary conditions to ensure the onset of the chiral edge states in the Kagom\'e magnet.  

\begin{table}
\caption{\label{table1}Atomistic material parameters and simulation parameters for T=0.001 K.}
\begin{tabular}{lcrc}
\hline
\hline
 1$^{\rm st}$ neighbours exchange interaction & $\mathcal{J}_{ij}$ & 1.0& (mRy)\\
Dzyaloshinskii-Moriya interaction & $|\mathcal{D}_{ij}|$ & 0.4 & (mRy)\\
Atomic Magnetic moment & s & 2.0 & ($\mu_{\rm B}$)\\
Damping & $\alpha_i$ & 0.001 & \\
Gyromatnetic ratio & $\gamma$ & 1.0 & ($\gamma_e$)\\
\hline
Global external field & $B_z$ & 0.2 & (T) \\
Simulation time & $t_s$ & 100 & (ps)\\
STM tip application time & $t_{STM}$ & 10 & (ps) \\
Local STM tip B field & $B_z$ & $-10^5$ & (T) \\
\hline
\hline
\end{tabular}
\end{table}

\begin{table}
\caption{\label{table2}Atomistic material parameters and simulation parameters for T=300 K. The exchange interaction has been varied in steps of 1 mRy while the DM interaction in steps of 0.4 mRy.}
\begin{tabular}{lcrc}
\hline
\hline
 1$^{\rm st}$ neighbours exchange interaction & $\mathcal{J}_{ij}$ & 1.0$\rightarrow$10.0& (mRy)\\
Dzyaloshinskii-Moriya interaction & $|\mathcal{D}_{ij}|$ & 0.4$\rightarrow$~4.0 & (mRy)\\
Atomic Magnetic moment & s & 2.0 & ($\mu_{\rm B}$)\\
Damping & $\alpha_i$ & 0.001 & \\
Gyromatnetic ratio & $\gamma$ & 1.0 & ($\gamma_e$)\\
\hline
Global external field & $B_z$ & 0.9 & (T) \\
Simulation time & $t_s$ & 100 & (ps)\\
STM tip application time & $t_{STM}$ & 10 & (ps) \\
Local STM tip B field & $B_z$ & $-10^7$ & (T) \\
\hline
\hline
\end{tabular}
\end{table}

\begin{figure}
\includegraphics[width=8.6 cm,angle=0]{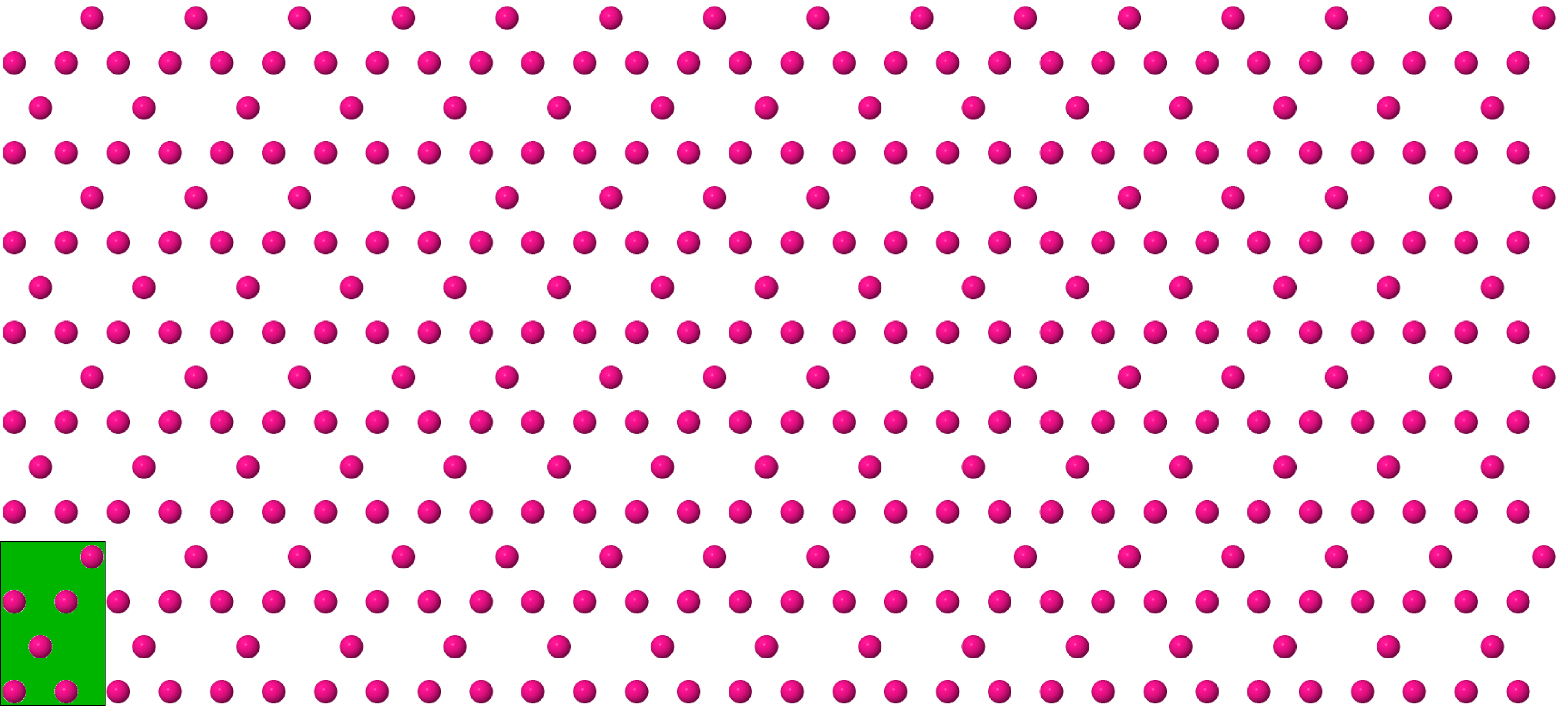}
\caption{\label{fig1}  {\bf Plot of a portion of the Kagom\'e stripe}. In green colour is plotted the unit cell. The total size of the calculated two-dimensional Kagom\'e lattice is $53\times400$ unit cells. 
}
\end{figure}

In Tables~\ref{table1} and~\ref{table2} we can observe that the intensity of the local magnetic field generated by the STM tip is rather huge. The reason for choosing those fields is just a technical detail related to the time we need for rendering the videos. In a ferromagnet, it is very well-known that the switching time (time required for reversing the spin direction) is very much dependent on the damping parameter \cite{1956JAP....27.1352K}. Thus, the switching time decreases as we increase the damping up to a critical value of the damping for which the switching time reaches a minimum value and then it continuously grows but in a more moderate way. In our simulations, the damping is very low ($\alpha$=0.001). This means that the switching time has to be quite large and in order to reduce this time to see the videos in a short period of time, we need to increase the magnetic field to an unrealistic value, but this technical detail does not affect at all the results of this work. In order to shed more light on this particular issue, we have performed several simulations with STM fields of 2000, 1000 and 100 T for four dampings ($\alpha=$0.005, 0.01, 0.05 and 0.08). In all cases for damping values less than 0.01 we have observed the skyrmion-antiskyrmion phase and above this damping the skyrmionic excitations are generated but they do not move and die after some time because of the dissipation process driven by the damping, as shown in Fig.~\ref{fig2} for a STM field of 1000 T. It is very clear how the skyrmion lifetime grows with the inverse of the damping parameter. Above 0.01 the skyrmions undergo a breathing behaviour characterized by a cyclic shrinking and enlargement of their size until eventually they collapse into a spin-wave. Below 0.01 the skyrmions are absolutely stable and they move just after their creation with the same properties as the ones we have generated with lower damping and larger local STM fields. The first message is that it is possible to reduce the values of the STM fields to more realistic values as 1 T or even less but waiting longer time to generate them, i.e., at least more than 0.1 ps. The second message is that for dampings in the interval from 0 to 0.01, the skyrmions can be succesfully stabilized so that the closer the damping is to 0.01, the lower the switching time becomes. A precise tuning of these parameters could give rise to an ``ultrafast'' generation of stable skyrmions.

\begin{figure}
\includegraphics[width=8.6 cm,angle=0]{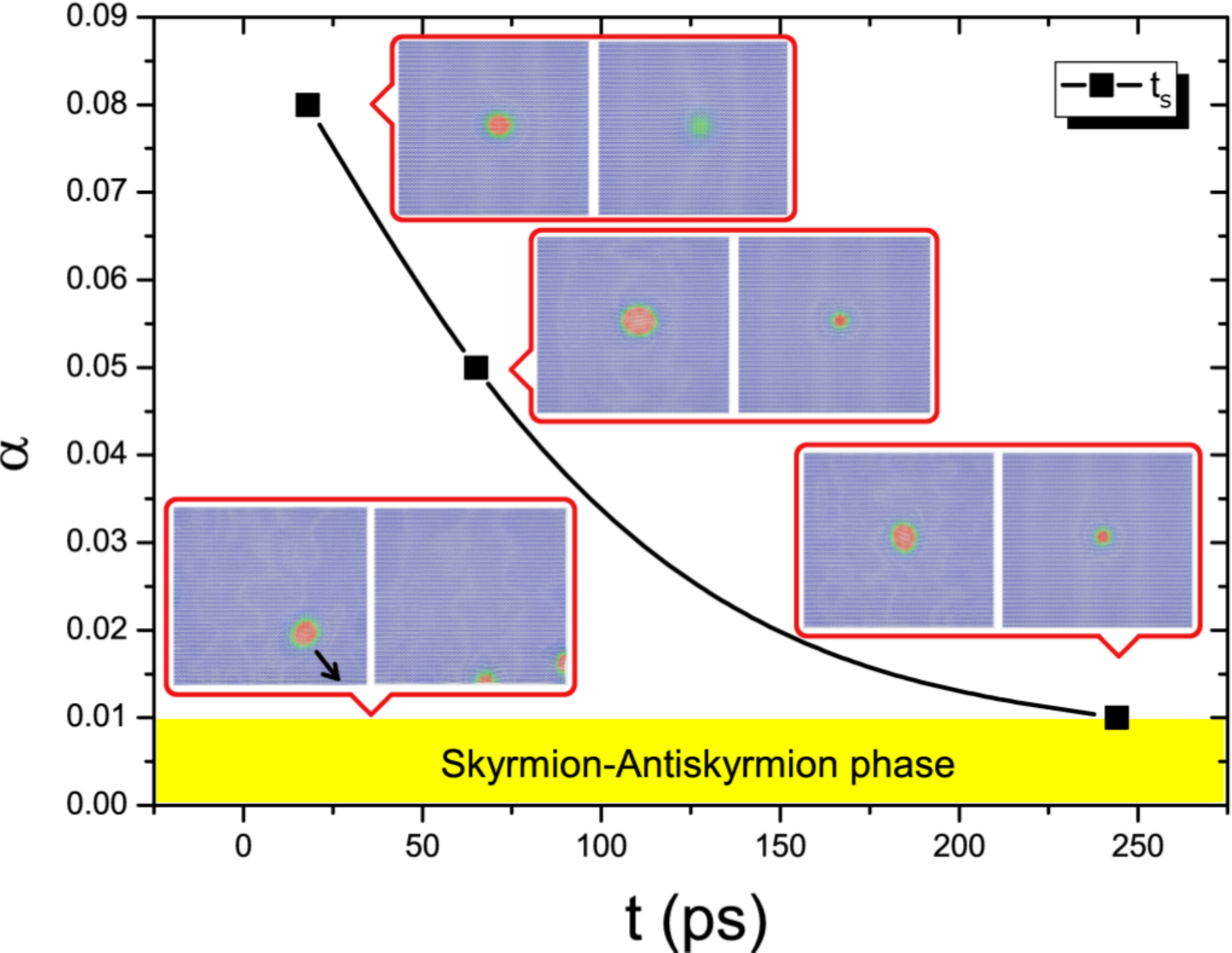}
\caption{\label{fig2}  {\bf Plot of damping parameter versus skyrmion lifetime $t_s$}. The figure is for an STM tip with a magnetic field of $10^3$ T. Some snapshots of skyrmions have been included in the figure in order to show how they reduce their size and eventually die. The topological excitations only survive in the skyrmion-antiskyrmion phase.
}
\end{figure}

\vspace{0.5cm}
\noindent{\bf S2: Calculation of the Berry curvature and topological invariants for a damped magnet in the tight-binding limit}
\vspace{0.5cm}

\noindent We start deriving the Berry curvature and the associated Euler characteristic assuming that we can grid the space into N different volume cells V$_j$. Every cell is represented by a monodomain with a pinned magnetic moment which is only allowed to rotate in space (tight-binding limit). The size of the volume cell not only can be at atomic level but also bigger or smaller. The magnetic moment or spin per unit cell is defined as
\begin{equation}
{\bf{s}}_j:=\left<\psi({\mathfrak{s}})\right|{\bf S}_j\left|\psi({\mathfrak{s}})\right>
\end{equation}   
where ${\bf S}_j=\int_{V_j}{\bf S}({\bf r}) d^3\bf{r}$ is the spin operator per unit cell and ${\bf{S}}({\bf{r}})$ represents the spin density operator. We denote $\mathcal{E}(\mathfrak{s}$) as the ground state energy for the spin configuration $\mathfrak{s}$ and $\psi(\mathfrak{s})$ is the corresponding wavefunction which satisfies the time-dependent Schr\"{o}dinger equation. In order to simplify the equations, we have intentionally removed the temporal dependency of the wavefunctions and quantum mechanical operators.

Hereafter, we used units in which $g \mu_\mathrm{B}=1$, i.e. $\gamma=\frac{1}{\hbar}$ where g is the g-factor,  $\mu_\mathrm{B}$ is the Bohr magneton and  ${\gamma}$ is the gyromagnetic ratio, respectively. Extremizing the action $\mathcal{S}=\int \mathcal{L} dt$ with respect to $\mathfrak{s}$ using the time-dependent variational principle we end up with the Euler-Lagrange equations or the equations of the spin dynamics. The Lagrangian is given by:

\begin{equation}
\mathcal{L}:=\left<\psi(\mathfrak{s})\right| i\hbar \frac{\partial}{\partial t}\left|\psi(\mathfrak{s})\right>-\left<\psi(\mathfrak{s})\right| \mathcal{H}\left|\psi(\mathfrak{s})\right>=\hbar\sum_{j=1}^\mathrm{N} \dot{{\bf s}}_j \mathcal{A}({\bf s}_j)-\mathcal{E} (\mathfrak{s})
\end{equation}
where $\mathcal{A}({\bf s}_j):=i\left<\psi(\mathfrak{s})\right|\frac{\partial}{\partial {\bf s}_j}\left|\psi(\mathfrak{s})\right>$ is the Berry connection. Energy dissipation can be implemented in the magnetic system adding a damped magnetization field to the equations of motion of the spin system

\begin{equation}
\frac{\partial\mathcal{R}(\dot{{\bf s}}_j)}{\partial \dot{{\bf s}}_j}=\eta_j \dot{{\bf s}}_j
\end{equation}
where $\mathcal{R}(\dot{{\bf s}}_j)$ is the local Rayleigh dissipation functional. This local functional allows us to define a site-resolved damping parameter $\eta_j$ in every single cell V$_j$. Taking into account that $\frac{\partial\mathcal{L}}{\partial \dot{{\bf s}}_j}=\hbar \mathcal{A}({\bf s}_j)$ and $\frac{\partial\mathcal{L}}{\partial {\bf s}_j}=\hbar\sum_{j=1}^\mathrm{N}\dot{{\bf s}}_j \frac{\partial  \mathcal{A}({\bf s}_j)}{\partial {\bf s}_j}-\frac{\partial \mathcal{E}(\mathfrak{s})}{\partial {\bf s}_j}$ is easy to show that the equations of motion are:
\begin{equation}
\hbar \sum_{j'=1}^\mathrm{N}\dot{{\bf s}}_{j'}\Theta({\bf s}_{j'},{\bf s}_j)-\eta_j \dot{{\bf s}}_j=-\frac{\partial \mathcal{E}(\mathfrak{s})}{\partial {\bf s}_j}
\label{equationofmotion}
\end{equation}
where the Berry curvature $\Theta({\bf s}_{j'},{\bf s}_j)$ is a 3x3 matrix. The ($\alpha$,$\beta$) element is given by:
\begin{equation}
\Theta_{j'j}^{\alpha\beta}:=\Theta(s_{j'}^\alpha,s_j^\beta)=\frac{\partial \mathcal{A}(s_{j'}^\alpha)}{\partial s_j^\beta}-\frac{\partial \mathcal{A}(s_j^\beta)}{\partial s_{j'}^\alpha}
\end{equation}
\vspace{0.5cm}
\noindent{\bf S2.1: Calculation of the Berry curvature in the rigid spin limit}
\vspace{0.5cm}

\noindent In the tight-binding limit, we rigidly rotate the spins within each unit cell, so that the constrained ground state $\psi(\mathfrak{s})$ is calculated as:
\begin{equation}
\left|\psi(\mathfrak{s})\right>=\prod_{j=1}^N \mathrm{e}^{i {\bm \theta}_j {\bf S}_j} \left|\psi(\mathfrak{s}_0)\right>
\end{equation} 
where $ {\bm \theta}_j\propto {\bf s}_j^0\times {\bf s}_j$ and $\psi(\mathfrak{s}_0)$ is the true ground state. Assuming for simplicity that ${\hat{z}}$ is the axis of symmetry, so that ${\bf s}_j^0= \mathrm{sgn}(s_j^0) s_j^0 \hat{z}$ and the norm of the spin is the same in every volume cell ($||{\bf s}_{j'}||=||{\bf s}_j||$), then after some cumbersome tensor algebra, the Berry curvature takes the form:

\begin{equation}
\Theta_{j'j}^{\alpha\beta}=-\frac{i}{{\bf s}_j^2}\left<\psi(\mathfrak{s}_0)\right|\left[ S_j^\alpha, S_{j'}^\beta\right]\left|\psi(\mathfrak{s}_0)\right>
\end{equation}  

Since that the spin operator ${\bf S}_j$ satisfies the commutation relations for the angular momentum operators, i. e. $[S_j^\alpha, S_{j'}^\beta]=i \delta_{jj'}\varepsilon^{\alpha\beta}_{\phantom{\alpha\beta}\gamma}  S_j^\gamma$, where $\varepsilon_{\alpha\beta\gamma}$ is the antisymmetric Levi-Civita symbol, the Berry curvature can be represented finally in terms of the spin components as\cite{1999PhRvL..83..207N}:

\begin{equation}
\Theta_{j'j}^{\alpha\beta}=\frac{1}{{\bf s}_j^2}\delta_{jj'} \varepsilon_{\alpha\beta\gamma}s_j^\gamma=\frac{\delta_{jj'}}{{\bf s}_j^2}\begin{pmatrix} 0 & s_j^z & -s_j^y \\ -s_j^z & 0 & s_j^x\\ s_j^y & -s_j^x & 0\end{pmatrix}
\label{berry_curvature}
\end{equation}
\vspace{0.5cm}
\noindent{\bf S2.2: Calculation of the Euler charateristic}
\vspace{0.5cm}

\noindent Let $\mathbb{S}$ be a manifold describing the parameter space of the spin fields. After the calculation of the Berry curvature, It is straightforward to calculate the Euler characteristic ($\chi(\mathbb{S})$) for a compact orientable manifold $\mathbb{S}$ using the generalized Gauss-Bonet theorem\cite{1999PhRvL..83..207N}, which states that:

\begin{equation}
(2\pi)^n\chi(\mathbb{S})=\int_\mathbb{S} e(\mathbb{S})=\int_\mathbb{S} \mathrm{Pf}(\Theta)
\label{eq_gauss}
\end{equation}
where e($\mathbb{S}$) is the Euler class and Pf($\Theta$) is the Pfaffian of the Berry curvature and $n$ is related to the dimension of the curvature. In a general case, the Berry curvature is calculated for an odd-dimensional manifold, i.e. the Berry curvature is a 3x3 matrix, so that the Pf($\Theta$)=0 and  consequently, Eq.~(\ref{eq_gauss}) gives $\chi(\mathbb{S})=0$.

In the particular case of a Kagom\'e lattice with the Hamiltonian given in the main text (Eq.~(1)) at very low damping ($\alpha$=0.001), the system has an axis of quantization along the $\hat{z}$ direction. The average in time, $<\cdots>_t$, gives $<s_j^z>_t=s_j^z$ and $<s_j^x>_t=<s_j^y>_t=0$. In consequence, only the x and y components of the Berry curvature survive. Then, for this system, $\Theta$ can be mapped into a $2\times2$ matrix:

\begin{equation}
<\Theta_{j'j}^{\alpha\beta}>_t=\frac{\delta_{jj'}}{{\bf s}_j^2}\begin{pmatrix}
0 & s_j^z\\-s_j^z&0 
\end{pmatrix}
\end{equation}
with $\alpha,\beta\in\{x,y\}$. Taking into account that Pf($\Theta$)=$s_j^z/{\bf s}_j^2$, $n$=1 and also that in a Kagom\'e lattice there are 3 atoms per unit cell, the Euler characteristic can be computed from the generalized Gauss-Bonet theorem:
\begin{equation}
\chi({\mathbb{S}})=\frac{3}{2\pi}\frac{\mathrm{sgn}(s_j^z)}{{\bf s}_j^2}\int_0^{s_j}\int_0^{s_j}\sqrt{{\bf s}_j^2-(s_j^x)^2-(s_j^y)^2} ds_j^xds_j^y
\end{equation}
The integral is easy to compute changing to spherical coordinates, i. e. $s_j^x=r_j \cos\theta_j$, $s_j^y=r_j \sin\theta_j$. Keeping in mind that the Jacobian determinant of this coordinate transformation is $\mathrm{det}(\mathcal{J})=\left|\frac{\partial (s_j^x,s_j^y)}{\partial(r_j,\theta_j)}\right|=r_j$, then $ds_j^xds_j^y=r_j dr_jd\theta_j$. Thus, the topological invariant takes the final form:
\begin{equation}
\chi({\mathbb{S}})=\frac{3}{2\pi}\frac{\mathrm{sgn}(s_j^z)}{{\bf s}_j^2}\int_0^{s_j}\int_0^{2\pi}\sqrt{{\bf s}_j^2-r_j^2} r_j dr_j d\theta_j=\mathrm{sgn}(s_j^z) s_j
\end{equation}
where $s_j$ is an integer number because of the bosonic nature of the magnonic excitations. For simplicity, we consider that $s_j=1$, then the Euler characteristic takes the values +1, 0 and -1. It means that in this system we have 3 different homotopical classes, so that they cannot be continuously deformed one into another. Likewise, every homotopical class is connected to a constrained ground state for the Kagom\'e system. The physical interpretation of this fact is as follows: Let's firstly reduce the problem to the unit cell of the Kagom\'e lattice, which has 3 atoms. If $\chi({\mathbb{S}})=+1$, then the three spins point to the +$\hat{z}$ direction while for $\chi({\mathbb{S}})=-1$ the spins are in the opposite direction. The third option is the case in
which $\chi({\mathbb{S}})=0$. To meet the requirement that $s_j^z$=0, the spins should lie in the Kagom\'e plane. In consequence, the system has a 3-fold degenerated state at T=0. It is interesting to note that a transition from a region, in spin space, with a positive Euler characteristic to a negative one is only possible through out a region with zero Euler characteristic because of the continuity in the spin rotation. It definitively opens the door to new topological magnetic solutions in the Kagom\'e lattice like skyrmions\cite{1961RSPSA.262..237S} or Belavin-Polyakov\cite{1975JETPL..22..245B} monopoles. 

\vspace{0.5cm}
\noindent {\bf S2.3: Spin dynamics equations}
\vspace{0.5cm}

\noindent In the tight-binding limit, the equations of motion can be reduced to the Landau$-$Lifshitz$-$Gilbert equations. To do so, the Berry curvature computed in Eq.~(\ref{berry_curvature}) is inserted in Eq.~(\ref{equationofmotion}):
\begin{equation}
\hbar \sum_{j'}^N \dot{s}_{j'}^\alpha \delta_{jj'} \varepsilon_{\alpha\beta\gamma}\frac{s_j^\gamma}{{\bf s}_j^2}+\frac{\partial \mathcal{E}(\mathfrak{s})}{\partial s_j^\alpha}-\eta_j \dot{s}_j^\alpha=0
\label{equation_landau}
\end{equation} 
Defining the effective field acting on the $j$ moment as $H_j^\alpha=\frac{\partial \mathcal{E}(\mathfrak{s})}{\partial s_j^\alpha}$, we can rewrite Eq.~(\ref{equation_landau}) in vectorial notation as:
\begin{equation}
-\hbar(\dot{{\bf s}}_j\times {\bf s}_j)+{\bf H}_j-\eta_j {\dot {\bf s}}_j=0
\end{equation}
with $\dot{{\bf s}}_j$ and ${\bf s}_j$ being unit vectors. After some algebra, we finally end up with the Landau$-$Lifshitz$-$Gilbert equations of motion for the spins:
\begin{equation}
\dot{{\bf s}}_j=\gamma({\bf s}_j\times {\bf H}_j)-\eta'_j({\bf s}_j\times \dot{{\bf s}}_j)
\end{equation} 
where $\eta'_j=\gamma \eta_j$ and $\gamma=\frac{1}{\hbar}$.

\end{document}